\documentclass[journal = jcisd8]{achemso}
\setkeys{acs}{articletitle = true}

\usepackage{amsfonts,amsthm,amssymb,amsmath,enumerate,graphicx,tikz,array, rotating}
\usepackage{todonotes}

\SectionNumbersOn

\newcommand{\NN}{\mathrm{NN}}
\newcommand{\m}{\mathrm{M}}
\newcommand{\mean}{\mathrm{mean}}

\theoremstyle{definition}

\title{Split Optimization for Protein/Ligand Binding Models}
\author{Brian Davis}
\affiliation{University of Kentucky - Markey Cancer Center}

\author{Kevin Mcloughlin}
\affiliation{Lawrence Livermore National Laboratory, Livermore, CA}

\author{Jonathan Allen}
\affiliation{Lawrence Livermore National Laboratory, Livermore, CA}

\author{Sally Ellingson}
\email{sally@kcr.uky.edu}
\affiliation{University of Kentucky - Markey Cancer Center}

\begin{document}
\maketitle
\begin{abstract}
    In this paper, we investigate potential biases in datasets used to make drug binding predictions using machine learning. We investigate a recently published metric called the Asymmetric Validation Embedding (AVE) bias which is used to quantify this bias and detect overfitting. We compare it to a slightly revised version and introduce a new weighted  metric. We find that the new metrics allow to quantify overfitting while not overly limiting training data and produce models with greater predictive value.
\end{abstract}

\section{Introduction}
Protein-ligand interactions are important to most processes in the human body, and therefore to regulating disease via drugs. There are an estimated 20,000 different human protein-coding genes\footnote{with about one-eighth of the exome containing observable genetic variations \cite{lek2016analysis}}, and $10^{60}$ small molecules in the  chemical universe\cite{reymond2012exploring}. Clearly, exploring all possible protein--drug pairs is not experimentally feasible. Drug discovery programs need accurate computational methods to predict protein--drug binding, and advances in machine learning have improved the accuracy of these predictions used in early stages of drug discovery. 
Area under the curve (AUC) scores for various curves are often provided to support suitability of a machine learning model for use in drug binding prediction--- in this paper we focus on the Precision--Recall curve and its associated AUC (PR-AUC).

The primary goal in protein--ligand binding modeling is to produce models that are capable of making accurate predictions on novel protein--drug pairs. Consequently, performance metrics need to reflect expected performance on novel data. This effort is frustrated by benchmark datasets that are not well-sampled from chemical space, so that novel pairs may be relatively far from data available to the modeler. Care must be taken in interpreting performance metrics like the PR-AUC, as laboratory experiments reveal that inferring real-world performance from AUC alone is overly-optimistic. This issue of generalizability is common in machine learning applications, but is particularly relevant in settings with insufficient and non-uniformly distributed data, as is the case with drug binding data.

The phenomenon of high performance metrics for low quality models is called overfitting, and is typically combated by using different data for the processes of model training and validation. If the validation data and real-world data for the model application are both distinct from the training data, then we expect the performance metrics on the validation data to be representative of the real-world performance of the model. A common way of splitting the available data into training and validation sets is to select a training ratio and randomly assign that proportion of the data to the training set.

A developing framework to account for overfitting is based on the assumption that the Nearest Neighbor (NN) model has poor generalizability. Within the context of reporting ``fair'' performance metrics, this working assumption of poor generalizability of NN models suggests several possibilities for more informative metrics, including:
\begin{enumerate}
    \item reporting the PR-AUC for a model produced from a training/validation split on which the Nearest Neighbor model has poor performance, and
    \item designing a metric which weights each validation molecule according to its relative distance to the binding classes in the training set. 
\end{enumerate}
We describe implementations of each of these approaches in this paper. For the first approach, we discuss the efforts presented in the Atomwise  paper\cite{atomwisePaper} to produce training/validation splits that are challenging for NN models, hereafter refered to as the Atomwise algorithm. We also describe two variations of the Atomwise algorithm: \mbox{ukySplit--AVE} and \mbox{ukySplit--VE}. As distinct from optimization, we introduce a weighting scheme $\omega$ designed to address the second approach, and discuss the consequences of using an $\omega$-weighted PR-AUC versus the traditional PR-AUC with a training/validation split produced by the \mbox{ukySplit--AVE} algorithm.

\subsection{Current Bias Quantification Methods}
Datasets with a metric feature space can be evaluated using spatial statistics \cite{rohrer2009maximum} to quantify the dataset topology and better understand potential biases. Of particular interest in the area of drug--binding model generalization are the ``nearest neighbor function'' G(t) and the ``empty space function'' F(t). G(t) is the proportion of active compounds for whom the distance to the nearest active neighbor is less than t. F(t) is the proportion of decoy compounds for whom the distance to the nearest active neighbor is less than t. Letting $\sum$G and $\sum$F denote the sum of the values of G and F over all thresholds t, it is reported that large values of $\sum$G indicate a high level of self-similarity and that small values of $\sum$F  indicate a high degree of separation. The difference of $\sum$G and $\sum$F gives a quick and interpretable summary of a dataset's spatial distribution, with negative values indicating clumping, near-zero values indicating a random-like distribution, and positive values indicating larger active-to-active distance than decoy-to-active. These spatial statistics were used to develop the Maximum Unbiased Validation (MUV) dataset, with the goal of addressing the reported association of dataset clumping with overly--optimistic virtual screening results  \cite{rohrer2009maximum,rohrer2008impact}. 

Wallach et al. \cite{atomwisePaper} extended the MUV metric, and used it to quantify the spatial distribution of actives and decoys among the training and validation sets.
 For two subsets $V$ and $T$ of a metric data set with distance function $d$, define, for each $v$ in $V$, the function $I_{t}(v,T)$ to be equal to one if $\min_{w\in T}\{d(v,w)\}<t$ and zero otherwise.
For a fixed value of $n$, define the function $H_{(V,T)}$ by
\begin{equation}\label{Eq:badDef}
H_{(V,T)}=\frac{1}{n+1}\cdot\frac{1}{|V|}\sum_{v\in V}\left(\sum_{i=0}^{n}I_{i/n}(v,T)\right).
\end{equation}
Then the Asymmetric Validation Embedding (AVE) bias is defined to be the quantity 
\begin{equation}\label{Eq:biasDef} B(V_A,V_I,T_A,T_I)=H_{(V_A,T_A)} - H_{(V_A,T_I)} + H_{(V_I,T_I)} - H_{(V_I,T_A)},
\end{equation} where the value of $n$ is taken to be 100, and where $V_A$ and $V_I$ are the validation actives and inactives (decoys), respectively, and similarly $T_A$ and $T_I$ are the training actives and inactives. For convenience, we abbreviate $H(V_a,T_a)-H(V_a,T_i)$ and $H(V_i,T_i)-H(V_i,T_a)$ as $(AA-AI)$ and $(II-IA)$, respectively. They are intended to be a quantification of the ``clumping'' of the active and decoy sets. If the term $(AA-AI)$ is negative, it suggests that, in the aggregate, the validation actives are closer to training decoys than to training actives, with the consequence that the active set is expected to be  challenging to classify. If the sum of $(AA-AI)$ and $(II-IA)$ (the AVE bias) is close to zero, it is expected that the data set is ``fair'', in that it does not allow for easy classification due to clumping. The authors also provide an AVE bias minimization algorithm. It is a genetic algorithm with breeding operations: merge, add molecule, remove molecule, and swap subset. The algorithm first generates initial subsets through random sampling, measures the bias, and selects subsets with low biases for breeding. The algorithm repeats bias scoring, redundancy removal, and breeding until termination based on minimal bias or maximum iterations.

In their paper, Wallach et al. observe that AUC scores\footnote{They report ROC-AUC scores, as opposed to PR-AUC scores.} and AVE bias scores are positively correlated for several benchmark data sets, implying that model performance is sensitive to the training/validation split.

In this paper, we present an efficient algorithm for minimizing the AVE bias of training/validation splits. We introduce a variation on the AVE bias, which we call the VE score, and describe its advantages in the context of optimization. We investigate the efficacy of minimizing these metrics for training/validation splits, and conclude by proposing a weighted performance metric as an alternative to the practice of optimizing training/validation splits.

\section{Methods}
\subsection{Dataset} Dekois 2 \cite{bauer2013evaluation} provides 81 benchmark datasets: 80 with unique proteins, and one with separate datasets for two different known binding pockets in the same protein. The active sets are extracted from BindingDB \cite{gilson2015bindingdb}. Weak binders are excluded, and 40 distinct actives are selected by clustering Morgan fingerprints by Tanimoto similarity. Three datasets are extended by selecting up to 5 actives from each structurally diverse cluster. The decoy set is generated using ZINC \cite{irwin2012zinc} and property matched to the actives based on molecular weight, octanol-water partition coefficient (logP), hydrogen bond acceptors and donors, number of rotatable bonds, positive charge, negative charge, and aromatic rings. Possible latent actives in the decoy set are removed using a score based on the Morgan fingerprint and the size of the matching substructures. Any decoy that contained a complete active structure as a substructure is also removed. 

\subsection{Bias metrics}
Throughout this paper, the term fingerprint refers to the 2048-bit Extended Connectivity Fingerprint (ECFP6) of a molecule as computed by the Python package RDKit \cite{RDKit}. 
For simplicity, we define 
\[d(v,T):=\min_{t\in T}\{d(v,t)\}\] and 
\[\Gamma(v,T):= \frac{\lfloor n\cdot d(v,T)\rfloor}{n+1},\]
where $d(v,t)$ is the Tanimoto distance between the fingerprints of the molecules $v$ and $t$.
We compute the AVE bias via the expression 
\begin{equation}\label{Eq:effBias}\mean_{v\in V_A}\{\Gamma(v,T_I)-\Gamma(v,T_A)\} + \mean_{v\in V_I}\{\Gamma(v,T_A)-\Gamma(v,T_I)\},
\end{equation} where $V_A$ and $V_I$ are the validation actives and inactives (decoys), respectively, and similarly $T_A$ and $T_I$ are the training actives and inactives. For a derivation of the equivalence of this expression and Expression (\ref{Eq:biasDef}), see the Appendix.
Since 
\[\left|d(v,T)-\Gamma(v,T)\right|<\frac{1}{n+1}
,\] for large values of $n$ Expression (\ref{Eq:effBias}) (and hence the AVE bias) is an approximation of 
\begin{equation}\label{Eq:Exact}\mean_{v\in V_A}\{d(v,T_I)-d(v,T_A)\} + \mean_{v\in V_I}\{d(v,T_A)-d(v,T_I)\}.
\end{equation} 
We now introduce the VE score, a close relative of the AVE bias:
\begin{equation}\label{Eq:VEscore}\sqrt{\mean^2_{v\in V_A}\{d(v,T_I)-d(v,T_A)\} + \mean^2_{v\in V_I}\{d(v,T_A)-d(v,T_I)\}}.
\end{equation}
While the raw ingredients of the AVE bias and the VE score are the same, they are qualitatively different, in particular as the VE score is never negative.



We generate a random training/validation split for each Dekois target and evaluate Expressions (\ref{Eq:biasDef}) through (\ref{Eq:VEscore}) 1,000 times with a single thread on an AMD Ryzen 7 2700x eight-core processor. We compare the mean computation times, as well as the computed values.

\subsection{Split Optimization}
We implement two custom genetic optimizers, ukySplit-AVE and ukySplit-VE, using the open source DEAP \cite{DEAP} framework. For a comparison of ukySplit-AVE and the Atomwise algorithm, see the Appendix. Both ukySplit-AVE and ukySplit-VE optimizers use parameters as described in Table \ref{Table:DEAPparams}. The parameters were chosen after grid-searching for minimum mean-time-to-debias on a sample of the Dekois targets.
\begin{table}
    \centering
    \begin{tabular}{|c|c|c|}\hline
        Parameter Name & Meaning & Value \\\hline
         POPSIZE&Size of the population& 500 \\\hline
         NUMGENS&Number of generations in the optimization& 2000\\\hline
         TOURNSIZE&Tournament Size & 4\\\hline
         CXPB&Probability of mating pairs&0.175\\\hline
         MUTPB&Probability of mutating individuals&0.4\\\hline
         INDPB&Probability of mutating bit of individual&0.005\\\hline
    \end{tabular}
    \caption{Evolutionary algorithm parameters}
    \label{Table:DEAPparams}
\end{table}

The optimizer populations consisted of training/validation splits, and the objective functions were given by Expressions (\ref{Eq:effBias}) and (\ref{Eq:VEscore}), respectively, for \emph{valid} splits, and equal to 2.0  otherwise. We say that a split is valid if
\begin{enumerate}
    \item the validation set contains at least one active and one decoy molecule,
    \item the active/decoy balance in the validation set is within $5\%$ of that in the total dataset,
    \item the ratio of training/validation set sizes is $80\pm 1\%$. 
\end{enumerate} 

\subsection{Modeling}
Using scikit-learn \cite{scikit-learn}, we train a random forest classifier ($n\_estimators=100$) with stratified 5-fold cross-validation and compute the mean PR-AUC for each target of the Dekois data set. We use fingerprints as features, and take the probability of the active class as the output of the model. For each of the folds, we evaluate the Expressions (\ref{Eq:effBias}) and (\ref{Eq:VEscore}), and report Pearson correlation coefficients with the PR-AUC. 

For the training/validation splits produced by an optimizer, we compute PR-AUC of a random forest model and evaluate Expression (\ref{Eq:effBias}) or  (\ref{Eq:VEscore}) as applicable.
 
\subsection{Nearest Neighbor similarity}
 We gather the binary predictions made by the Nearest Neighbor model, which predicts the class of a validation molecule to be the same as its nearest neighbor (using the metric $d$) in the training set. Considering the $\NN$ predictions as a bit string, we can compare it with the prediction bit string of any other model $\m$ using the Tanimoto similarity $T$: 
\[T(\NN, \m) = \; \frac{\sum \left(\NN \wedge \m\right) }{\sum \left(\NN \vee \m\right)},\] with bitwise operations $\wedge$ (and) and $\vee$ (or) and sums over all predictions for the validation set. We take the maximum Tanimoto similarity over all thresholds $\eta$ for each of the validation folds, and report the mean.

\subsection{Weighted PR-AUC} For a given model, let TP, TN, FP, and FN be the collections of molecules for which the model predictions are true positive, true negative, false positive, and false negative, respectively. The metrics precision and recall may be easily generalized by assigning a weight $\omega(v)$ to each molecule $v$, and letting the $\omega$--weighted precision be given by
\[\frac{\sum_{v\in \text{TP}}\omega(v)}{\sum_{v\in \text{TP}}\omega(v) + \sum_{v\in \text{FP}}\omega(v)}\]
and the $\omega$--weighted recall be given by
\[\frac{\sum_{v\in \text{TP}}\omega(v)}{\sum_{v\in \text{TP}}\omega(v) + \sum_{v\in \text{FN}}\omega(v)}.\]
Setting the weight $\omega(v)$ equal to 1 for all molecules $v$, we recover the standard definitions of precision and recall. 

Inspired by Expression (\ref{Eq:Exact}), we define the ratio $\gamma(v)$ by 
\[\gamma(v)=\begin{cases} & \frac{d(v, T_A)}{d(v,T_I)}\quad \text{ if $v$ is active,}\\
& \frac{d(v, T_I)}{d(v,T_A)}\quad \text{ if $v$ is decoy.}\end{cases}\]

When we refer to the weighted PR-AUC in this paper we use the weight $\omega$ given by the cumulative distribution function of $\gamma$ over the validation set for the target protein. Note that the weights $\omega$ are between zero and one, and that the weighting de-emphasizes molecules that are closer to training molecules of the same binding class than to training molecules of the opposite class. Thus the larger the contribution of a molecule to the AVE bias, the lower its weight. For further description of the $\omega$--weighted PR-AUC, see the Appendix.

\subsection{Generalization Ability}
Inspired by recent work presented on the so-called ``far AUC''\cite{farAUC}, we attempt to measure the ability of a drug-binding model to generalize. We randomly split the data set for each target 80/20 (preserving the class balance), then remove any molecules in the 80\% set that are distance less than 0.4 from the 20\% set. We reserve the 20\% set to serve as a proxy for novel data ``far'' from the training data. We then treat the remainder of the 80\% as a data set, running the same analysis as described in the earlier subsections: computing the weighted and un-weighted PR-AUC of a random forest trained on random splits, as well as the PR-AUC of random forest models trained on ukySplit-AVE and ukySplit-VE optimized splits.

\section{Results}
\subsection{Computational Efficiency}\label{sec:efficient_evaluation}
A naive implementation of Equation (\ref{Eq:biasDef}) required a mean computation time over all Dekois targets of 7.14 ms, while an implementation of Equation (\ref{Eq:effBias}) had a mean computation time of 0.99 ms. The mean computation times for Expressions (\ref{Eq:Exact}) and (\ref{Eq:VEscore}) were both approximately 0.31 ms. 

\begin{figure}
\includegraphics[scale=0.5]{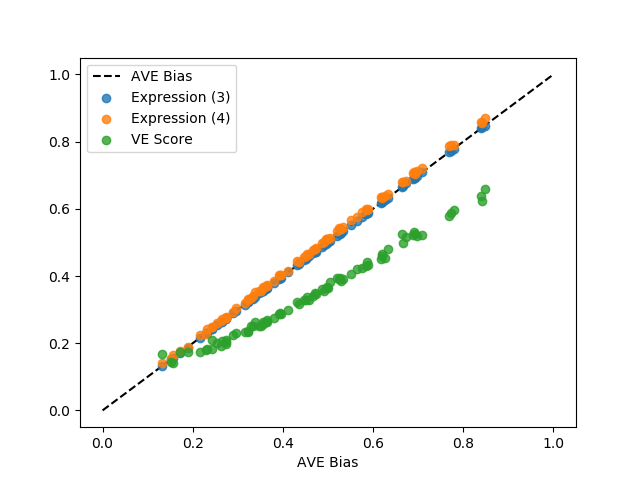}
\caption{Comparison of evaluations over Dekois targets}
\label{Fig:newScoreAcc}
\end{figure}

\begin{table}
    \centering
    \begin{tabular}{|c|c|c|}\hline
         Expression & Mean Computation Time & Relative Speedup  \\\hline
         (\ref{Eq:biasDef})& 7.14 ms&1\\\hline
         (\ref{Eq:effBias})& 0.99 ms & 7.2\\\hline
         (\ref{Eq:Exact})& 0.31 ms&23.4\\\hline
         (\ref{Eq:VEscore})& 0.31 ms&23.1\\\hline
    \end{tabular}
    \caption{Computational Efficiency}
    \label{Table:Speedups}
\end{table}
Evaluations of Expressions (\ref{Eq:biasDef}) through (\ref{Eq:VEscore}) are plotted in Figure \ref{Fig:newScoreAcc}. The absolute differences between the computed value of Expression (\ref{Eq:biasDef}) and Expressions (\ref{Eq:effBias}) and   (\ref{Eq:Exact}) are summarized in Table \ref{Table:differences}. It is not meaningful to compare the evaluations of Expressions (\ref{Eq:biasDef}) and (\ref{Eq:VEscore}) in this way, as they measure different, though related, quantities. 
\begin{table}
    \centering
    \begin{tabular}{|c|c|c|}\hline
         Expression&Mean Abs Difference&Max Abs Difference  \\\hline
        (\ref{Eq:effBias}) & $3.1 \times 10^{-4}$ & $4.1 \times 10^{-3}$\\\hline
        (\ref{Eq:Exact}) & $9.9 \times 10^{-3}$ & $2.3 \times 10^{-2}$\\\hline
    \end{tabular}
    \caption{Comparison with Expression (\ref{Eq:biasDef}) over Dekois targets}
    \label{Table:differences}
\end{table}

For reference, the AVE paper considered a score of $2 \times 10^{-2}$ to be ``bias free''.

\subsection{Split Bias and Model Performance}

\begin{figure}[ht]
\includegraphics[scale=0.5]{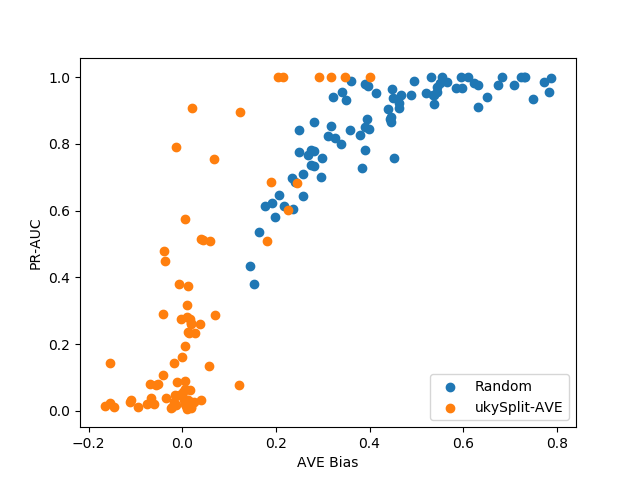}
\caption{Mean Split AVE Bias vs. Model PR-AUC.}
\label{Fig:AVE_PRAUC}
\end{figure}

In Figure \ref{Fig:AVE_PRAUC}, we plot the mean PR-AUC against the mean AVE bias for 5-fold cross validation on each Dekois target. The Pearson correlation coefficient between mean AVE bias and mean PR-AUC is computed to be 0.80, which is comparable in strength to the correlation reported in the AVE paper for other benchmark datasets. We also plot the AVE bias against the mean PR-AUC for each target after optimization by ukySplit--AVE. Note that, although the optimizer was run with a stopping criterion of 0.02, it is possible for the minimum AVE bias to jump from greater than 0.02 to a negative number (as low as -0.2) in one generation.

\begin{figure}[ht]
\includegraphics[scale=0.5]{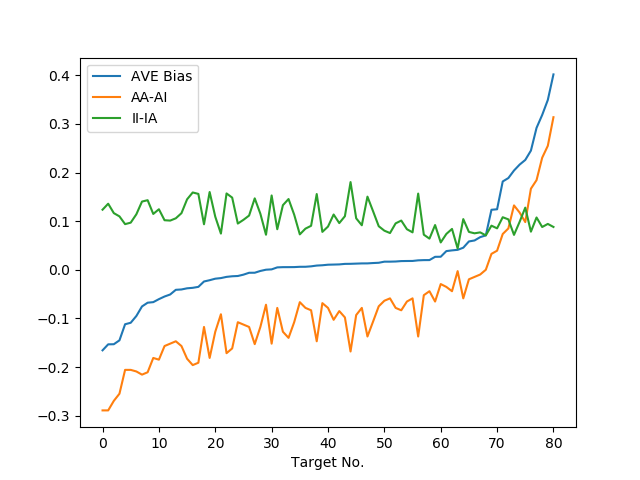}
\caption{The two components of the AVE Bias score.}
\label{Fig:atomwise_score_components}
\end{figure}

We order the target proteins by AVE bias, and plot the two components, AA-AI and II-IA, after optimization by ukySplit--AVE in Figure \ref{Fig:atomwise_score_components}.

\subsection{Optimization by ukySplit--VE}

\begin{figure}[ht]
\includegraphics[scale=0.5]{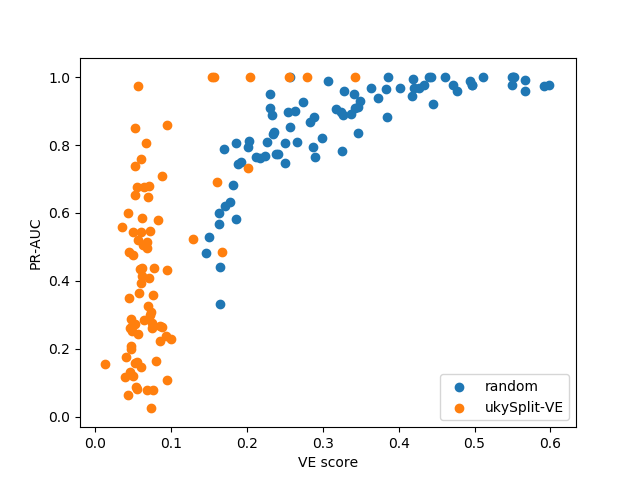}
\caption{Mean VE Score vs. Model PR-AUC.}
\label{Fig:VE_PRAUC}
\end{figure}

\begin{figure}[ht]
\includegraphics[scale=0.5]{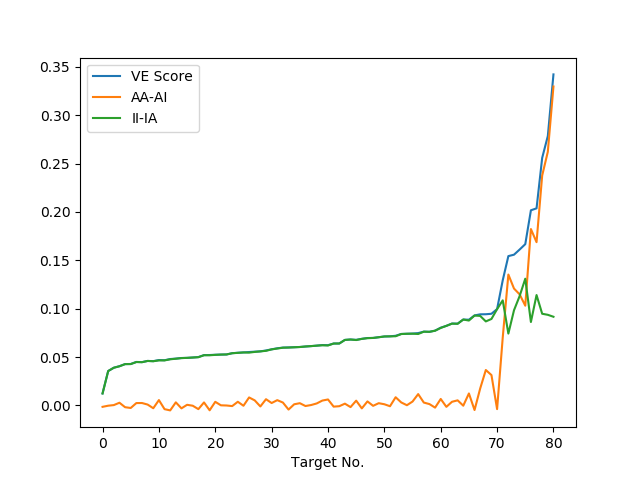}
\caption{The two components of the VE Score.}
\label{Fig:new_score_components}
\end{figure}

Figure \ref{Fig:VE_PRAUC} plots the mean VE score against the mean PR-AUC across each fold of the cross-validation split for each target before and after optimization with ukySplit--VE (minimizing VE score as opposed to AVE bias). Figure \ref{Fig:new_score_components} plots the score components associated to the active and decoy validation molecules after optimizing VE score (for comparison with Figure \ref{Fig:atomwise_score_components}).

\subsection{Weighted Performance}
\begin{figure}[ht]
\includegraphics[scale=0.5]{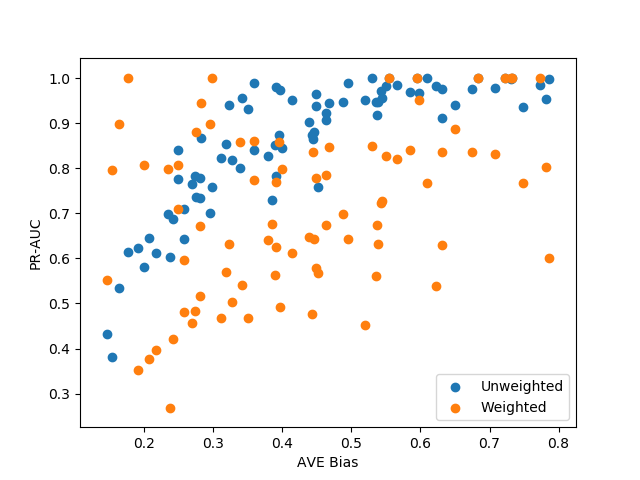}
\caption{Mean Split AVE Bias vs. Model Weighted PR-AUC.}
\label{Fig:weighted_PRAUC}
\end{figure}

Figure \ref{Fig:weighted_PRAUC} plots the $\omega$--weighted PR-AUC against the mean AVE bias over each fold for each target. Recall that the models and predictions are the same as those represented in Figure \ref{Fig:AVE_PRAUC} with label ``random'', but that the contributions of each validation molecule to the precision and recall are now weighted by the weight function $\omega$.

\subsection{Nearest Neighbor similarity}
\begin{figure}[ht]
\includegraphics[scale=0.5]{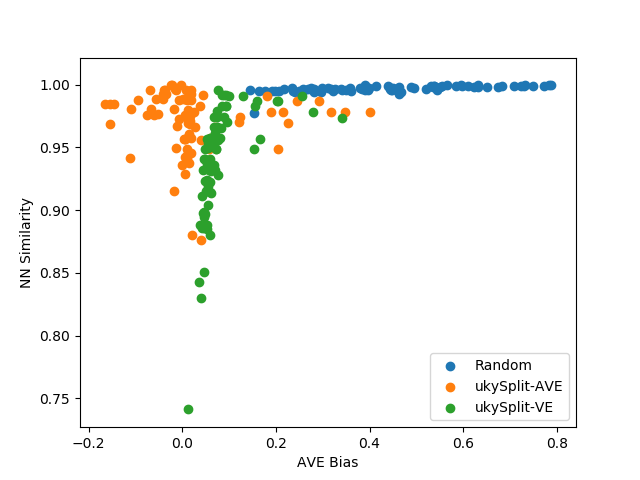}
\caption{NN similarity of a random forest model trained on various splits.}
\label{Fig:NN_comparison}
\end{figure}

Figure \ref{Fig:NN_comparison} plots the NN- similarity of a random forest model trained on splits produced randomly, by ukySplit-AVE, and by ukySplit-VE. The mean NN-similarities were 0.997, 0.971, and 0.940, respectively.

\subsection{Model Performance on Distant Data}
After reserving 20\% of each target's data for the test set, approximately 3\% of the remainder was found to be within the 0.4 buffer distance, and was removed before splitting into training and validation sets. 
Similarly to Sundar and Colwell \cite{farAUC}, we find that de-biasing training/validation splits does not lead to increased performance on ``distant'' test sets: the mean ratio of test PR-AUC before and after split optimization by ukySplit-AVE was 1.010 (1.018 for ukySplit-VE).

\begin{figure}[ht]
\includegraphics[scale=0.75]{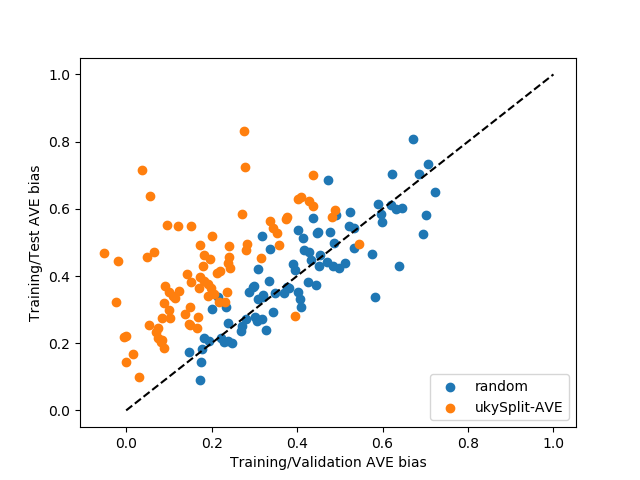}
\caption{AVE bias of training/validation and training/test splits.}
\label{fig:validation_test_AVE}
\end{figure}

\begin{figure}[ht]
\includegraphics[scale=0.75]{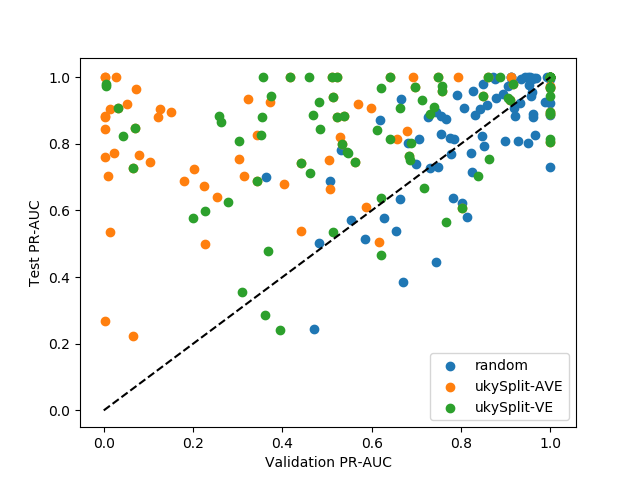}
\caption{Model performance on validation vs. test set of a model trained on various splits.}
\label{fig:farAUC}
\end{figure}

Figure \ref{fig:validation_test_AVE} plots the AVE bias on the training/test split against the AVE bias on the training/validation split (letting the test set play the role of the validation set in the AVE bias definition).
Figure \ref{fig:farAUC} shows the validation and test PR-AUC of a model trained with a training set produced randomly, by ukySplit-AVE, and by ukySplit-VE. 

\section{Discussion}
\subsection{Computing Bias}
As presented in Table \ref{Table:Speedups}, refactoring Expression (\ref{Eq:biasDef}) into Expression (\ref{Eq:effBias}) yielded speedups of 7x, and the additional use of exact, rather than approximated, values yielded a speedup of roughly 23x for Expression (\ref{Eq:Exact}). While Expressions (\ref{Eq:biasDef}) and (\ref{Eq:effBias}) are mathematically equivalent, in practice they yield slightly different results due to machine precision. In the aggregate over the Dekois dataset, the difference is negligible relative to the established definition of ``de-biased'', as described in Table \ref{Table:differences}. Expressions (\ref{Eq:biasDef}) and (\ref{Eq:Exact}) are \emph{not} mathematically equivalent. In light of the equivalence of Expressions (\ref{Eq:biasDef}) and (\ref{Eq:effBias}), it is clear that AVE bias (Expression (\ref{Eq:biasDef})) is an \emph{approximation} of Expression (\ref{Eq:Exact}). Their difference, though slight, is properly interpreted as approximation error in the AVE bias.

\subsection{Model effects of optimization}\label{sec:pathologicalSplits}
Figures \ref{Fig:AVE_PRAUC} and \ref{Fig:VE_PRAUC} demonstrate that the process of minimizing bias in the training / validation split risks training a model with little or no predictive value. The expected recall (and precision) for a random guessing model is equal to the balance of active molecules, which for the Dekois dataset is 3\%. Of the 81 Dekois targets, 21 (about 26\%) had models with below random PR-AUC when trained and validated on a split produced by ukySplit--AVE. This may be understood by considering Figure \ref{Fig:atomwise_score_components}, which shows that the AVE bias is primarily an indication of the spatial distribution of the (minority) active class in the validation set, with low AVE bias associated with active validation molecules that are closer to training decoys than to training actives. Models trained on such splits are therefore prone to misclassify validation actives, and hence have a low PR-AUC. This phenomenon is less pronounced when splits are optimized for VE score (ukySplit--VE), as it does not allow terms to ``cancel out'', and so does not incentivize pathological distributions of validation actives. Only one Dekois target had worse-than-random performance for a model trained on a split optimized for VE score, and while the mean PR-AUC over models trained with ukySplit--AVE splits was 0.26, the mean PR-AUC for models trained on ukySplit--VE splits was 0.44.

As demonstrated in Figure \ref{Fig:NN_comparison}, random forest models trained on ukySplit-AVE optimized splits are still substantially similar to Nearest Neighbor models.  

\subsection{Weighted PR-AUC}
\begin{figure}
\includegraphics[scale=0.5]{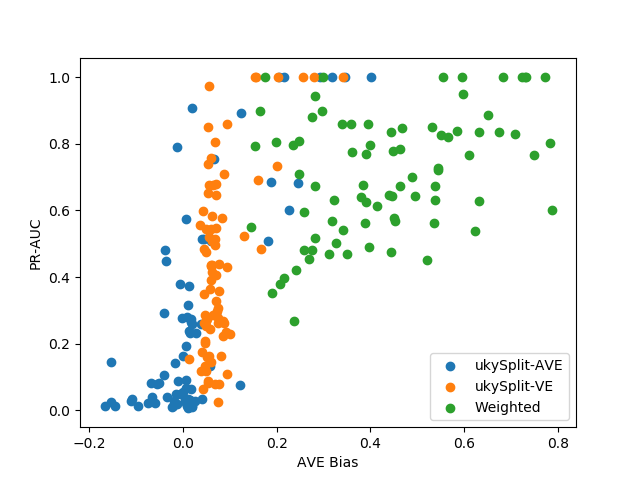}
\caption{Comparison of ukySplit-AVE, ukySplit-VE, and weighting methods.}
\label{Fig:comparison}
\end{figure}

As described in the Introduction, a weighted metric represents an alternative way of taking into account the assumption of poor generalizability of Nearest Neighbor models. While bias optimization creates training/validation splits that are more challenging for Nearest Neighbor models, they simultaneously result in low quality models, even when using powerful methods like random forests. In contrast, the weighted metric $\omega$-PR-AUC discounts the potentially inflated performance of models without degrading the models themselves (see Figure \ref{Fig:comparison}). It is worth noting, as well, that the computational expense of computing the weighting $\omega$ is negligible compared with the intensive work performed in optimizing training / validation splits.

Ultimately, we don't claim that that the weighted PR-AUC is a good indicator of expected model performance on novel data. Rather, it may be used to infer that the standard PR-AUC is inflated due to the spatial distribution of the validation set. In particular, if two trained models have the same performance, the one with the higher weighted performance may be expected to have better generalizability.



\subsection{Test Performance}
Figure \ref{fig:validation_test_AVE} shows that minimizing the AVE bias on the training/validation split does not minimize the AVE bias on the training/test split.
Figure \ref{fig:farAUC} demonstrates that even when a split results in a trained model with very low validation PR-AUC, the model still performs fairly well on the test data.

\section{Conclusions}

Developers of machine learning models for virtual high-throughput screening will have to contend with issues of over-fitting as long as drug binding data is scarce. While optimization of training/validation splits may have merit, the practice of minimizing AVE bias is not a valid method for addressing this issue for several reasons:
\begin{enumerate}
    \item the resulting models tend to have low predictive value,
    \item minimizing the AVE bias of a training/validation split does not seem to effect the AVE bias of the training/test split, and
    \item the resulting models are still essentially equivalent to the nearest-neighbor model, and are not expected to generalize well.
\end{enumerate}

We propose the use of weighted performance metrics as a less computation-intensive alternative to split optimization. We do not claim that the weighted metrics are more representative of expected performance on novel data, but rather as a warning that overfitting may be taking place. If the weighted and un-weighted metrics diverge, we can conclude that the good performance of a model is concentrated at data points on which a nearest-neighbor model is sufficient.

\newpage
\bibliography{main}
\bibliographystyle{plain}

\newpage
\section{Appendix}
\subsection{AVE bias}
There is room for improvement in Equation (\ref{Eq:badDef}):
\begin{equation*}
H_{(V,T)}=\frac{1}{n+1}\cdot\frac{1}{|V|}\sum_{v\in V}\left(\sum_{i=0}^{n}I_{i/n}(v,T)\right).
\end{equation*}
Notice that $d<i/n$ if and only if $nd<i$. There are two cases to consider. First, if $nd$ is an integer, then $nd<i$ if and only if $nd\leq i-1$. Second, if $nd$ is not an integer, then $nd<i$ if and only if $\lceil nd\rceil\leq i$. In both cases, $\sum_{i=0}^nI_{i/n}(v,T)$ is equal to the number of $i$ satisfying $0\leq i\leq n$ and $nd<i$, and in both cases this number is $n-\lfloor nd\rfloor$.

For simplicity, we define 
\[d(v,T):=\min_{t\in T}\{d(v,t)\}\] and \[\Gamma(v,T):= \frac{\lfloor n\cdot d(v,T)\rfloor}{n+1}.
\]
Then 
\[B(V_A,V_I,T_A,T_I) = \mean_{v\in V_A}\{\Gamma(v,T_I)-\Gamma(v,T_A)\} + \mean_{v\in V_I}\{\Gamma(v,T_A)-\Gamma(v,T_I)\},
\] as presented in Equation (\ref{Eq:effBias}).

\subsection{Atomwise algorithm vs ukySplit-AVE}
Of the 81 Dekois targets, ukySplit-AVE successfully removed the bias from 59, while Atomwise removed the bias from 26.

\begin{table}
\caption{Breakdown of successes for Atomwise and ukySplit-AVE methods.}
\begin{tabular}{|c|c|c|}\hline
     &ukySplit-AVE Successful&ukySplit-AVE Unsuccessful  \\\hline
     Atomwise Successful& 26 & 0\\\hline
     Atomwise Unsuccessful & 33& 22\\\hline
\end{tabular}
\end{table}

Among the targets for which neither method was successful in removing bias, the mean final AVE bias scores were 0.15 and 0.24, respectively.

\subsection{Weighted PR-AUC}
\begin{figure}[ht]
\includegraphics[scale=0.5]{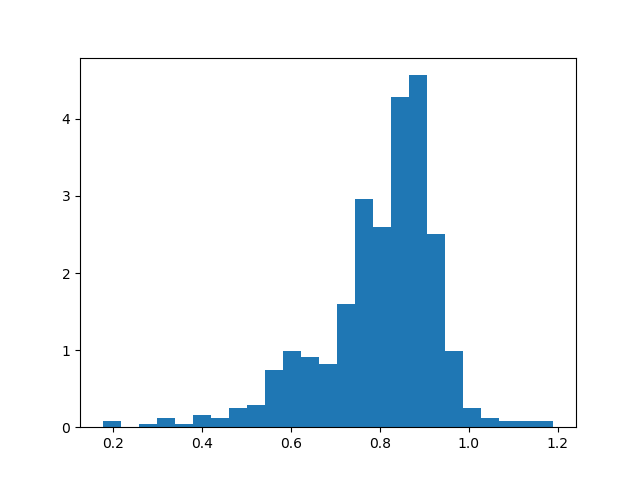}
\caption{The distribution of $\gamma$.}
\label{fig:cox1_prob}
\end{figure}

\begin{figure}[ht]
\includegraphics[scale=0.5]{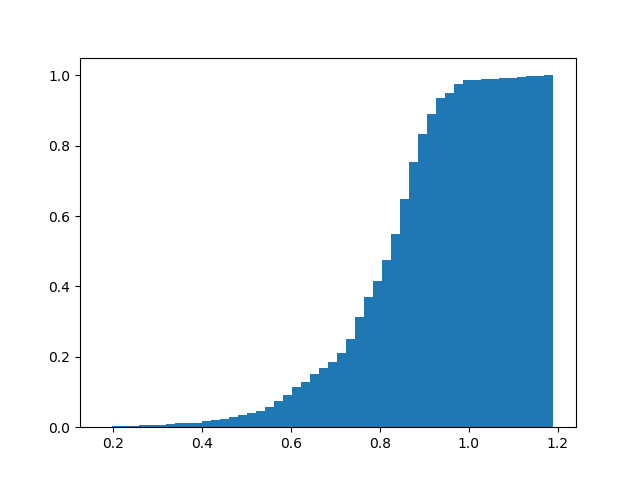}
\caption{The cumulative distribution of $\gamma$.}
\label{fig:cox1_cum}
\end{figure}

We give more details of the weighting scheme $\omega$ with the example of Dekois target protein COX1. We generate a valid random split and for purposes of illustration choose an active molecule $v$ at random from the validation set. The distance to the nearest active in the training set is 0.47, while the distance to the nearest decoy in the training set is 0.81. The ratio $\gamma(v)$ is then given by $0.47/0.81 = 0.58$. Note that $\gamma$ is less than one, indicating that this validation molecule is closer to the  training molecules of the same binding class than the opposite binding class. The contribution of this molecule to the AVE bias of this split is approximately $(0.81-0.47)/23=0.015$ (using Expression \ref{Eq:Exact}).
The distribution of the values of $\gamma$ over the entire validation set is presented in Figure \ref{fig:cox1_prob}, and the corresponding cumulative distribution is presented in Figure \ref{fig:cox1_cum}.

The weight $\omega(v)$ for our example validation molecule is equal to 0.09, the evaluation of the cumulative distribution function at $\gamma(v)=0.58$. It may also be interpreted as the percentile (9\%) of the value of $\gamma(v)$ among all $\gamma$ values.

For a random forest model trained on this particular random split, the predicted probability that $v$ is active is 0.4. Using a threshold value of 0.3, the resulting prediction is that $v$ is active (a true positive). With standard weights ($\omega=1$), the confusion matrix for this threshold is given by Table \ref{table:confusion}, but with the weighting $\omega$ the table is given by Table \ref{Table:weightedConfusion}.

\begin{table}
    \centering
    \begin{tabular}{|c|c|c|}\hline
        &Pred Active&Pred Decoy\\\hline
         Active&7&16 \\\hline
         Decoy&2&575\\\hline
    \end{tabular}
    \caption{Confusion matrix for threshold 0.3}
    \label{table:confusion}
\end{table}

\begin{table}
    \centering
    \begin{tabular}{|c|c|c|}\hline
        &Pred Active&Pred Decoy\\\hline
         Active&3.21&13.43 \\\hline
         Decoy&1.25&331.65\\\hline
    \end{tabular}
    \caption{$\omega$-weighted Confusion matrix for threshold 0.3}
    \label{Table:weightedConfusion}
\end{table}

The resulting precision and recall are 0.78 and 0.3, respectively, while the $\omega$-weighted precision and recall are .72 and 0.19, respectively. The decrease in these metrics indicates that the molecules correctly classified by the model have below average weight, i.e., the spatial distribution of the validation set is artificially inflating the performance metrics.

Repeating this calculation for threshold values between 0 and 1.0 yields the $\omega$--weighted precision recall curve, and we call the area under the curve the $\omega$-PR-AUC.

\end{document}